\documentclass[preprint]{aastex}

\usepackage{hyperref}
\usepackage{lscape}

\begin{document}

\title{The Galactic Magnetic Field's Effect in Star-Forming Regions}

\author{Ian W. Stephens\altaffilmark{1}, Leslie W. Looney\altaffilmark{1}, C. Darren Dowell\altaffilmark{2}, John E. Vaillancourt\altaffilmark{3,4}, Konstantinos Tassis\altaffilmark{2}}

\altaffiltext{1}{Department of Astronomy, University of Illinois at Urbana-Champaign, 1002 West Green Street, Urbana, IL 61801, USA; stephen6@illinois.edu}
\altaffiltext{2}{Jet Propulsion Laboratory, California Institute of Technology, Pasadena, CA 91109, USA}
\altaffiltext{3}{Division of Physics, Mathematics, \& Astronomy, California Institute of Technology, Pasadena, CA, 91125 USA}
\altaffiltext{4}{Current address: Stratospheric Observatory for Infrared Astronomy, Universities Space Research Association, NASA Ames Research Center, MS 211-3, Moffet Field, CA 94035, USA}


\begin{abstract}

  We investigate the effect of the Milky Way's magnetic field in star
  forming regions using archived 350\,$\mu$m polarization data on 52
  Galactic star formation regions from the Hertz polarimeter module. The polarization
  angles and percentages for individual telescope beams were combined
  in order to produce a large-scale average for each source and for
  complexes of sources.  In more than 80\% of the sources, we find
  a meaningful mean magnetic field direction, implying the
  existence of an ordered magnetic field component at the scale of
  these sources.  The average polarization angles were analyzed with
  respect to the Galactic coordinates in order to test for correlations
  between polarization percentage, polarization angle, intensity, and
  Galactic location.  No correlation was found, which suggests that
  the magnetic field in dense molecular clouds is decoupled from the
  large-scale Galactic magnetic field.  Finally, we show that the
  magnetic field directions in the complexes are consistent with a
  random distribution on the sky.

\end{abstract}

\section{INTRODUCTION}

Magnetic fields are known to play an important role in star formation
\citep[e.g.,][]{mc99,mck07}. The line-of-sight component of the
magnetic field is often measured using the Zeeman effect or Faraday
rotation \citep[e.g.,][]{cht03}.  The angle of the magnetic field with respect to the
plane of the sky can be deduced through continuum polarization
measurements. Far-infrared continuum polarization is due to emission from elongated
dust grains that align perpendicularly to the magnetic field
\citep[e.g.,][]{hil88}. For the same reason, optical light that has undergone 
partial absorption by dust grains exhibits polarization
parallel to the magnetic field \citep[e.g.,][]{dav51}.

In general the magnetic field lines of the Milky Way follow the
direction of the spiral arms \citep[often measured via Faraday
rotation of pulsar signals; e.g.,][]{han06}.  Since dust polarization
vectors are orthogonal to magnetic field lines, the Galactic magnetic field
 should induce a tendency for mapped polarization angles to be perpendicular to
the Galactic disk. External galaxies also show magnetic fields
following the spiral arms \citep[e.g.,][]{sof86}, but on small scales
(about 20\,pc in the case of NGC 6946) localized processes (e.g., star
formation) dominate, tangling magnetic field lines \citep{bec07}.

The 350\,$\mu$m Hertz polarimeter module was located at the Caltech
Submillimeter Observatory and operated from 1994 to 2003. 
The Hertz polarimeter was well-suited for polarization measurements that probe the 
dense environments around forming stars, specifically clumps and cores.
Since the decommissioning of Hertz,
\citet{dot10} have published an archive of 56
different objects, 52 of which are Galactic star
forming regions. 
In this paper, we use this relatively large dataset of star-forming regions
to calculate a single large-scale 
(up to an angular diameter of about 10 arcminutes) average degree of polarization ($P$, percentage of wave that is polarized), position angle ($\theta$), and flux density ($I$) for
each of the 52 star-forming Hertz datasets. The data are then explored
to test whether the magnetic fields of these regions have any statistical
relationship to the Galactic magnetic field.


\citet{eri96} applied a similar analysis on visible polarization of
3000 stars (from the \citet{mat70} dataset) that are in the Galactic
plane. They found that the measured optical polarization angles from most
stars, particularly those at further distances, are parallel to the
Galactic magnetic field. \citet{fos02} found that optical polarization
measurements had a sinusoidal dependence with respect to Galactic
longitude. 

However, we note that the observations probe very different physical 
regions: submillimeter observations
probe dense regions ($A_V \gtrsim 30$) while optical observations
probe diffuse regions ($A_V \lesssim 5$). 
Although dust polarization of star formation regions has been compared
to the Galactic plane magnetic field before, the studies have been limited to
small sample sizes.
Continuum polarimetry at $\lambda$~=~0.8 and 1.3 mm of approximately 10 star formation
sources (at scales of $\sim$1\arcmin) 
revealed no correlation for the magnetic field direction with respect to the Galactic plane
\citep{gle99}.
On the other hand, \citet{li06} analyzed the dust polarization of 4 giant
molecular clouds (capturing most of the clouds on scales of $\sim$10\arcmin)
at 450\,$\mu$m and found that 3 of these have a significant
field aligned within $15^\circ$ of the Galactic plane. Our dataset
probes similar regions, but we have a much larger sample size
that will constrain, with much higher confidence, the relationship of the magnetic field to the Galactic plane.

Our results do not show any similar correlation for degree of polarization 
or angle with Galactic location.   An in-depth analysis was
applied for $\theta$ by binning data based on Galactic longitude and
spiral arm locations, again suggesting no correlation.

This paper is organized as follows: in \S \ref{oda} we discuss
our data and the analysis methods used. It also shows that a single polarization angle for each dataset is typically meaningful.
Our results are presented in \S \ref{results} and discussed in \S \ref{discussion}.

\section{Observational Data and Analysis}\label{oda}

For the 52 analyzed Hertz datasets\footnote{Data available online
  \citep{dot10}.}, there were an average of about 90 telescope beams
per object. For each telescope beam, the following values were of
interest: degree of polarization ($P$), polarization position angle
($\theta$), flux density ($I$), and the uncertainties on each. Outliers were removed for measurements with $\sigma_\theta\geq 76^\circ$, which is the $2\sigma$ point of a Gaussian distribution with a $90^\circ$ full width at half maximum (most datasets have this or better). After these cuts, the average and median telescope beams per cloud were 82 and 56 respectively.

For each object, $P$, $\theta$, and $I$ values were used to calculate
the Stokes parameters $Q$ and $U$. These were then summed to find
the large-scale average, converting back to mean $P$s and
$\theta$s. Uncertainties were propagated. A similar technique was
used in \citet{tas09}.  The mean angles were rotated from equatorial
to Galactic coordinates. The spherical trigonometry for this
transformation is discussed in the appendix.



\subsection{Stokes Parameters, $P$s, and $\theta$s}

Since we are not able to measure circular polarization with Hertz, the
Stokes parameter $V$ is assumed to be 0; it is expected to actually be
near this value. Circular polarization is also not relevant for average
field direction. From $I_i$, $P_i$, and $\theta_i$, measured at a beam location ($i$)
in a source map, the following formulas
were used to find the corresponding $Q_i$ and $U_i$.

\begin{equation}
	Q_i = I_iP_i\cos2\theta_i
\end{equation}
\begin{equation}
	U_i = I_iP_i\sin2\theta_i
\end{equation}

From here, unweighted sums of $Q_i$ and $U_i$ were calculated in
order to obtain $Q_\mathrm{tot} = \sum Q_i$ and $U_\mathrm{tot} = \sum
U_i$. Similarly, an unweighted sum for intensity, $I_\mathrm{tot} =
\sum I_i$, was found. These quantities are not weighted by standard deviation and 
represent the sums as if the clouds were observed with a larger telescope 
beam.


By using error propagation to calculate $\sigma_{Q_i}$ and $\sigma_{U_i}$, it is found:

\begin{equation}
 \sigma_{Q_i}^2 = (P_i\sigma_{I_i} \cos2\theta_i)^2 + (I_i\sigma_{P_i} \cos2\theta_i)^2 +
(2I_iP_i\sigma_{\theta_i} \sin 2\theta_i)^2
\end{equation}
\begin{equation}
 \sigma_{U_i}^2 = (P_i\sigma_{I_i} \sin2\theta_i)^2 + (I_i\sigma_{P_i} \sin2\theta_i)^2 +
(2I_iP_i\sigma_{\theta_i} \cos 2\theta_i)^2
\end{equation}

The variances for $Q_\mathrm{tot}$ and $U_\mathrm{tot}$ were found by summing the variances of each beam measurement:

\begin{equation}
  \sigma_{Q_\mathrm{tot}}^2 =\sum_{i=1}^{n}\sigma_{Q,i}^2
\end{equation}

This method of calculating the standard deviation accounts for errors due to measurement
in uncorrelated linear datasets. The variance for $I_\mathrm{tot}$ was calculated similarly. 

 From $Q_\mathrm{tot}$, $U_\mathrm{tot}$, and $I_\mathrm{tot}$, $\langle P\rangle$ and $\langle\theta\rangle$ can now be calculated. $\langle P\rangle$ is seen below:

\begin{equation}
	\langle P\rangle = \frac{\sqrt{(Q_\mathrm{tot})^2+(U_\mathrm{tot})^2}}{I_\mathrm{tot}}
\end{equation}

The calculation of $\langle\theta\rangle$ is more complicated. $\langle\theta\rangle$ must always be positive, and the arctangent term does not discriminate on which Qs and Us are negative. The following formulae successfully determines the proper $\langle\theta\rangle$.

 \begin{equation}
   \langle\theta\rangle' =  \left\{
     \begin{array}{lr}
       0.5\arctan \frac{U_\mathrm{tot}}{Q_\mathrm{tot}}  & :
\arctan\frac{U_\mathrm{tot}}{Q_\mathrm{tot}} \geq 0\\
       0.5\left(\pi + \arctan\frac{U_\mathrm{tot}}{Q_\mathrm{tot}}\right)  & :
\arctan\frac{U_\mathrm{tot}}{Q_\mathrm{tot}} <0
     \end{array}
   \right.
\end{equation} 


 \begin{equation}
   \langle\theta\rangle =  \left\{
     \begin{array}{lr}
       \langle\theta\rangle'  & :
U_\mathrm{tot} \geq 0\\
      \langle\theta\rangle' + \pi/2  & :
U_\mathrm{tot} <0
     \end{array}
   \right.
\end{equation} 

In order to calculate the variances for $\langle P\rangle$ and $\langle\theta\rangle$, propagation of error was used again:

\begin{equation}
  \sigma_{\langle P\rangle}^2 = \frac{Q_\mathrm{tot}^2\sigma_{Q_\mathrm{tot}}^2+U_\mathrm{tot}^2\sigma_{U_\mathrm{tot}}^2}{I_\mathrm{tot}^2(Q_\mathrm{tot}^2+U_\mathrm{tot}^2)}+
\frac{(Q_\mathrm{tot}^2+U_\mathrm{tot}^2)\sigma_{I_\mathrm{tot}}^2}{I_\mathrm{tot}^4}
\end{equation}


\begin{equation}
\sigma_{\langle\theta\rangle}^2 = \frac{Q_\mathrm{tot}^2\sigma_{U_\mathrm{tot}}^2+U_\mathrm{tot}^2\sigma_{Q_\mathrm{tot}}^2}{4(Q_\mathrm{tot}^2+U_\mathrm{tot}^2)^2}
\end{equation}

It is important to note that equations (3), (4), (9), and (10) do not incorporate the non-diagonal (correlation) terms of covariance matrices, which are necessary for proper propagation of error. The \citet{dot10} datasets do not report correlation terms from their original coordinate transformation of $I$, $Q$, and $U$ to $P$ and $\theta$. Additionally, other correlation terms exist, such as observational correlations and correlations between adjacent beams.
More accurate calculations of uncertainties are not motivated for the analysis and conclusions of this paper, but it should be noted that without the correlation terms, uncertainties are approximate.

Our results for a large-scale average are consistent with the polarimetric maps in \citet{dot10}. 
Although de-biasing $\langle P\rangle$, as discussed by \citet{val06}, is an important part in the analysis of $\langle P\rangle$ implied by $\sigma_{\langle P\rangle}$, it was not done on these data since it does not modify the polarization angle (the primary focus for results in this paper).

The Hertz averages were analyzed in three ways: 1) investigating each of
the 52 Hertz sources separately; 2) separating the sources by Galactic
arm; and 3) combining near-by databases (i.e., data within $5^\circ$
in Galactic longitude and less than 200\,pc apart along the line of
sight). The last technique reduces the amount of total datasets to 22 
``complexes.''

\subsection{Mean Angle Significance}

To investigate whether a meaningful mean direction of the magnetic field exists or whether, instead, 
angles in the Hertz datasets are random (follow a uniform angle distribution) we used a Kolmogorov-Smirnov (KS) 
test. 
Out of the 52 datasets, 42 were inconsistent with
a random distribution of angles at a 90\% confidence level. By looking
at the polarimetric maps in \citeauthor{dot10}, the 10 datasets that exhibited 
a scatter consistent with a 
random distribution were maps that had poor signal-to-noise or had
``circular-like'' morphology (e.g., W75N). 
In this paper, our analysis was conducted in three ways: keeping these ten datasets; omitting these datasets; and analyzing these datasets by themselves. In all cases, 
we arrived at consistent conclusions. 

Similarly, for the 22 complexes, 17 were inconsistent with a random distribution of angles at a 90\% confidence level.  Thus, the data have a meaningful large-scale average angle that is 
inconsistent with random processes for $>$80\% of the sources and $>$77\% of the complexes
with 90\% confidence.

A similar conclusion can also be obtained by using the
normalized Stokes parameters\footnote{The primes denote the difference between the typical definition of normalized Stokes parameters, i.e., $q=Q/I$ and $u=U/I$.}:
\begin{equation}
q^\prime  = \cos 2\theta\,,
\end{equation}
\begin{equation}
u^\prime = \sin 2\theta\,.
\end{equation}

The values of the normalized Stokes parameters $q^\prime$ and $u^\prime$ at each
pointing position within a cloud must be in the range
[$-1$,$+1$]. Random samples drawn from a uniform distribution of
angles should have an average $q^\prime$ and $u^\prime$ of 0.  With a finite number
of samples we can only expect that average to reach zero within some
uncertainty.  With approximately 56 samples per cloud, from a uniform
distribution of angles we expect $\langle q^\prime \rangle$ approximately $0
\pm 0.09$.  For the Hertz sample of 52 clouds, $\langle q^\prime \rangle$ and
$\langle u^\prime \rangle$ are in the range $-0.90$ to 0.74, significantly
above the expectations for a uniform distribution of angles.

When taking the averages of the magnitudes of $\langle q^\prime \rangle$ and
$\langle u^\prime \rangle$ for the clouds and complexes, a similar argument
holds. We define $q_\mathrm{cloud} = \langle |\langle q^\prime \rangle|\rangle$ and
$u_\mathrm{cloud} = \langle |\langle u^\prime \rangle|\rangle$ (i.e., the average
magnitudes of $\langle q^\prime \rangle$ and $\langle u^\prime \rangle$) for the 52
clouds. The uncertainty due to the finite sample of 52 clouds is
slightly increased from the previous paragraph. That is, for a uniform
random distribution of angles we expect $q_\mathrm{cloud}$ and $u_\mathrm{cloud}$ to
have a standard deviation of 0.10.  The actual values for our dataset are
$q_\mathrm{cloud} = 0.284$ and $u_\mathrm{cloud} = 0.322$, significantly larger than
what could be expected from a uniform random distribution.  For the 22
complexes a uniform distribution gives a standard deviation of 0.15. The
actual values for our dataset are $q_\mathrm{complex} = 0.196$ and
$u_\mathrm{complex} = 0.226$, which are both significant at an 80\%
confidence level.

\section{Results}\label{results}

In this section we present an extensive examination
of possible correlations between Galactic coordinates ($l$ and $b$),
degree of polarization ($\langle P \rangle$), Galactic polarization angle ($\langle\theta_G\rangle$; transformation from $\langle\theta\rangle$ is discussed in the appendix), and intensity ($\langle I \rangle$). Section 3.1 focuses on correlations with Galactic coordinates. Section 3.2 bins data based on Galactic arms. In both cases no relationship between location and $\langle\theta_G\rangle$ was established. Table 1 shows the data for each individual object and its associated complex; note that large clouds (e.g., OMC-1 for the fourth complex in Table 1) may be dominant for the calculated Galactic angle for a complex. Table 2 summarizes the coefficient of determination (i.e., the square of the Pearson product-moment correlation coefficient), $R^2$, for most of the comparisons done. Finally, Section 3.3 examines correlation orthogonal to the Milky Way's magnetic field.
\begin{deluxetable}{lcccccccccccccccc}
\tabletypesize{\scriptsize}
\tablewidth{0pt}
\rotate
\tablehead{\colhead{Source$^a$} & \colhead{$\alpha$ (2000)} & \colhead{$\delta$ (2000)} & \colhead{$l$} & \colhead{$b$} & \colhead{$\langle I\rangle$} & \colhead{$\sigma_{\langle I\rangle}$} & \colhead{$\langle P\rangle$} & \colhead{$\sigma_{\langle P\rangle}$} & \colhead{$\langle\theta\rangle$} & \colhead{$\langle\theta_G\rangle$} & \colhead{$\sigma_{\langle\theta\rangle}$} & \colhead{$\langle\theta_G\rangle_{co}^b$} & \colhead{$\sigma_{\langle\theta\rangle_{co}}^b$} & \colhead{Spiral} & \colhead{Dist.$^c$} \\ & \colhead{(h:m:s)} & \colhead{(d:m:s)} & \colhead{(deg)} & \colhead{(deg)} & \colhead{(Jy)} & \colhead{(Jy)} & \colhead{(\%)} & \colhead{(\%)} & \colhead{(deg)} & \colhead{(deg)} & \colhead{(deg)} & \colhead{(deg)} & \colhead{(deg)} & \colhead{Arm} & \colhead{(kpc)}}
\startdata
W3$^1$ & 02:25:40.7 & 62:05:52 & 133.71 & 1.22 & 142.1 & 1.3 & 0.69 & 0.04 & 71.4 & 50.9 & 1.8 & 50.9 & 1.8 & Perseus & 1.95$\pm$0.04(1)\\
NGC 1333$^2$ & 03:29:03.7 & 31:16:03 & 158.35 & -20.56 & 19.8 & 0.3 & 0.36 & 0.43 & 84.9 & 48.1 & 34.3 & 48.1 & 34.3 & Local & 0.318$\pm$0.027(2)\\
L1551$^3$ & 04:31:34.2 & 18:08:05 & 178.93 & -20.05 & 17.2 & 1.6 & 1.23 & 0.62 & 41.6 & 170.9 & 14.3 & 170.9 & 14.3 & Local & 0.14$\pm$0.01(3)\\
IRAS 05327-0457$^4$ & 05:35:14.4 & -04:57:38 & 208.57 & -19.18 & 19.3 & 0.2 & 3.02 & 0.28 & 143.3 & 80.6 & 2.1 & 143.6 & 0.2 & Local & 0.4(4)\\
OMC-1$^4$ & 05:35:14.5 & -05:22:32 & 208.99 & -19.38 & 213.7 & 1.1 & 1.86 & 0.01 & 27.6 & 144.8 & 0.2 &   &   & Local & 0.4(4)\\
OMC-2$^4$ & 05:35:26.7 & -05:10:00 & 208.82 & -19.25 & 56.8 & 0.7 & 0.54 & 0.05 & 138.8 & 76.1 & 2.7 &   &   & Local & 0.4(4)\\
OMC-3$^4$ & 05:35:23.5 & -05:01:32 & 208.68 & -19.19 & 42.5 & 0.1 & 1.72 & 0.07 & 135.6 & 72.9 & 1.3 &   &   & Local & 0.4(4)\\
OMC-4$^4$ & 05:35:08.2 & -05:35:56 & 209.19 & -19.51 & 24.0 & 2.7 & 1.13 & 0.17 & 167.7 & 104.8 & 4.3 &   &   & Local & 0.4(4)\\
L1641N$^4$ & 05:36:18.8 & -06:22:11 & 210.06 & -19.59 & 24.2 & 1.9 & 0.66 & 0.24 & 36.3 & 153.1 & 10.1 &   &   & Local & 0.4(4)\\
NGC 2023$^5$ & 05:41:25.4 & -02:18:06 & 206.86 & -16.60 & 19.1 & 3.8 & 1.30 & 0.35 & 129.1 & 67.0 & 7.8 & 84.4 & 1.6 & Local & 0.4(4)\\
NGC 2024$^5$ & 05:41:43.0 & -01:54:22 & 206.53 & -16.36 & 174.2 & 2.4 & 0.51 & 0.03 & 154.8 & 92.8 & 1.8 &   &   & Local & 0.4(5)\\
HH24MMS$^5$ & 05:46:08.4 & -00:10:43 & 205.49 & -14.57 & 10.8 & 1.2 & 1.19 & 0.45 & 90.5 & 28.8 & 10.8 &   &   & Local & 0.4(5)\\
NGC 2068 LBS 17$^5$ & 05:46:28.0 & -00:00:54 & 205.38 & -14.42 & 8.2 & 0.4 & 0.76 & 0.35 & 49.0 & 167.4 & 13.2 &   &   & Local & 0.4(5)\\
NGC 2068 LBS 10$^5$ & 05:46:50.2 & 00:02:01 & 205.38 & -14.32 & 16.8 & 0.4 & 3.24 & 0.14 & 128.2 & 66.5 & 1.2 &   &   & Local & 0.4(5)\\
NGC 2071$^5$ & 05:47:04.8 & 00:21:47 & 205.11 & -14.11 & 44.1 & 1.0 & 0.46 & 0.07 & 147.4 & 85.9 & 4.6 &   &   & Local & 0.4(5)\\
Mon R2$^6$ & 06:07:46.6 & 06:23:16 & 213.71 & -12.60 & 140.8 & 3.7 & 0.59 & 0.04 & 30.7 & 147.4 & 1.8 & 164.9 & 2.5 & Local & 0.83$\pm$.05(6)\\
GGD12$^6$ & 06:10:50.4 & -06:11:46 & 213.88 & -11.84 & 95.0 & 3.6 & 0.86 & 0.07 & 90.8 & 27.6 & 2.2 &   &   & Local & 1(7)\\
S269$^7$ & 06:14:36.6 & 13:49:35 & 196.45 & -1.68 & 18.8 & 2.1 & 1.76 & 0.46 & 37.1 & 155.9 & 7.4 & 155.9 & 7.4 & Perseus & 3.8(8)\\
AFGL 961$^8$ & 06:34:37.7 & 04:12:44 & 207.27 & -1.81 & 10.7 & 1.0 & 1.14 & 0.36 & 160.9 & 98.6 & 9.0 & 98.6 & 9.0 & Perseus & 1.7(9)\\
Mon OB1 27$^9$ & 06:40:58.3 & 10:36:54 & 202.30 & 2.53 & 5.8 & 0.7 & 0.52 & 0.46 & 157.0 & 94.4 & 24.9 & 56.4 & 3.7 & Local & 0.8(10)\\
Mon OB1 25$^9$ & 06:41:03.7 & 10:15:07 & 202.63 & 2.38 & 16.5 & 4.1 & 1.93 & 0.67 & 116.8 & 54.2 & 9.8 &   &   & Local & 0.8(10)\\
Mon OB1 12$^9$ & 06:41:06.1 & 09:34:09 & 203.24 & 2.08 & 21.6 & 0.1 & 1.42 & 0.14 & 138.7 & 76.1 & 2.9 &   &   & Local & 0.8(10)\\
NGC 2264$^9$ & 06:41:10.3 & 09:29:27 & 203.32 & 2.06 & 64.6 & 1.1 & 0.57 & 0.04 & 81.1 & 18.6 & 2.2 &   &   & Local & 0.8(10)\\
$\rho$ Oph$^{10}$ & 16:26:27.5 & -24:23:54 & 353.08 & 16.91 & 46.6 & 4.0 & 1.24 & 0.07 & 161.0 & 29.4 & 1.6 & 27.1 & 2.2 & Local & 0.139$\pm$0.006(11)\\
IRAS 16293-2422$^{10}$ & 16:32:22.8 & -24:28:36 & 353.94 & 15.84 & 33.9 & 0.1 & 0.41 & 0.12 & 90.4 & 139.6 & 8.1 &  &  & Local & 0.178$^{+0.018}_{-0.037}$(12)\\
CB68$^{11}$ & 16:57:19.5 & 16:09:21 & 4.50 & 16.34 & 3.8 & 0.5 & 0.51 & 0.34 & 55.9 & 110.2 & 19.1 & 110.2 & 19.1 & Local & 0.16(13)\\
NGC 6334V$^{12}$ & 17:19:57.4 & -35:57:46 & 351.16 & 0.70 & 193.9 & 7.0 & 0.21 & 0.05 & 105.9 & 160.7 & 6.2 & 100.0 & 1.1 & Sagittarius & 1.74$\pm$0.31(14)\\
NGC 6334A$^{12}$ & 17:20:19.1 & -35:54:45 & 351.25 & 0.67 & 187.0 & 2.8 & 1.19 & 0.04 & 68.8 & 123.6 & 1.1 &   &   & Sagittarius & 1.74$\pm$0.31(14)\\
NGC 6334I$^{12}$ & 17:20:53.4 & -35:47:00 & 351.42 & 0.65 & 487.8 & 7.2 & 0.70 & 0.03 & 36.5 & 91.4 & 1.2 &   &   & Sagittarius & 1.74$\pm$0.31(14)\\
M-0.13-0.08$^{13}$ & 17:45:37.3 & -29:05:40 & 359.87 & -0.08 & 156.5 & 3.6 & 0.86 & 0.06 & 107.1 & 165.4 & 2.0 & 140.9 & 0.9 & Gal.\ center & 8(15)\\
Sgr A East$^{13}$ & 17:45:41.5 & -29:00:09 & 359.95 & -0.05 & 132.3 & 1.2 & 1.12 & 0.03 & 92.1 & 150.4 & 0.8 &   &   & Gal.\ center & 8(15)\\
CO 000.02-00.02$^{13}$ & 17:45:42.1 & -28:56:05 & 0.01 & -0.01 & 139.7 & 5.8 & 1.08 & 0.03 & 53.5 & 111.8 & 0.7 &   &   & Gal.\ center & 8(15)\\
M-0.02-0.07$^{13}$ & 17:45:51.6 & -28:59:09 & 359.99 & -0.07 & 127.0 & 2.8 & 1.96 & 0.10 & 80.9 & 139.3 & 1.4 &   &   & Gal.\ center & 8(15)\\
M+0.07-0.08$^{13}$ & 17:46:04.3 & -28:54:45 & 0.07 & -0.07 & 82.3 & 3.0 & 0.58 & 0.08 & 37.5 & 95.8 & 3.9 &   &   & Gal.\ center & 8(15)\\
M+0.11-0.08$^{13}$ & 17:46:10.2 & -28:53:06 & 0.11 & -0.08 & 144.8 & 4.6 & 0.09 & 0.05 & 132.2 & 10.5 & 16.9 &   &   & Gal.\ center & 8(15)\\
M+0.25+0.01$^{13}$ & 17:46:10.5 & -28:42:17 & 0.26 & 0.02 & 135.9 & 11.8 & 0.36 & 0.06 & 93.5 & 151.9 & 4.7 &   &   & Gal.\ center & 8(15)\\
M+0.34+0.06$^{13}$ & 17:46:13.2 & -28:36:53 & 0.34 & 0.05 & 87.5 & 3.8 & 1.02 & 0.11 & 42.5 & 100.9 & 3.1 &   &   & Gal.\ center & 8(15)\\
Sickle (G0.18-0.04)$^{13}$ & 17:46:14.9 & -28:48:03 & 0.19 & -0.05 & 99.9 & 9.7 & 1.48 & 0.23 & 114.8 & 173.2 & 4.5 &   &   & Gal.\ center & 8(15)\\
M+0.40+0.04$^{13}$ & 17:46:21.4 & -28:35:41 & 0.38 & 0.04 & 75.2 & 2.1 & 1.11 & 0.13 & 160.2 & 38.6 & 3.4 &   &   & Gal.\ center & 8(15)$^d$\\
Sgr B1$^{13}$ & 17:46:47.2 & -28:32:00 & 0.48 & -0.01 & 172.5 & 4.1 & 1.00 & 0.04 & 128.4 & 6.9 & 1.1 &   &   & Gal.\ center & 8(15)\\
Sgr B2$^{13}$ & 17:47:20.2 & -28:23:06 & 0.67 & -0.04 & 1013.2 & 3.5 & 0.36 & 0.01 & 77.1 & 135.6 & 1.0 &   &   & Gal.\ center & 8(15)\\
W33 C (G12.8-0.2)$^{14}$ & 18:14:13.4 & -17:55:32 & 12.81 & -0.20 & 206.3 & 2.3 & 0.21 & 0.03 & 41.7 & 102.9 & 4.3 & 114.0 & 26.2 & Scutum-Crux$^e$ & 4.5(16)\\
W33 A$^{14}$ & 18:14:38.9 & -17:52:04 & 12.91 & -0.26 & 83.5 & 1.7 & 0.57 & 0.07 & 129.1 & 10.3 & 3.4 &   &   & Scutum-Crux$^e$ & 4.5(16)\\
L483$^{15}$ & 18:17:29.8 & -04:39:38 & 24.88 & 5.38 & 9.2 & 0.3 & 0.32 & 0.19 & 163.3 & 45.1 & 16.8 & 45.1 & 16.8 & Local & 0.2(17)\\
M 17$^{16}$ & 18:20:24.5 & -16:13:02 & 15.01 & -0.69 & 276.3 & 5.6 & 0.82 & 0.02 & 164.0 & 45.6 & 0.8 & 45.6 & 0.8 & Sagittarius & 1.6$^{+0.3}_{-0.1}$(18)\\
W43-MM1$^{17}$ & 18:47:46.9 & -01:54:29 & 30.82 & -0.06 & 154.3 & 4.4 & 0.80 & 0.17 & 178.9 & 61.4 & 6.1 & 61.4 & 6.1 & Scutum-Crux$^f$ & 7.0$\pm$0.9(19)\\
G34.3+0.2$^{18}$ & 18:53:18.5 & 01:14:59 & 34.26 & 0.15 & 206.6 & 15.6 & 0.51 & 0.16 & 109.3 & 171.9 & 8.8 & 171.9 & 8.8 & Scutum-Crux$^g$ & 3.7(20)\\
W49 A$^{19}$ & 19:10:13.6 & 09:06:17 & 43.17 & 0.01 & 153.1 & 5.7 & 0.45 & 0.03 & 56.7 & 119.0 & 2.0 & 119.0 & 2.0 & Perseus & 11.4$\pm$1.2(21)\\
W51 A (G49.5-0.4)$^{20}$ & 19:23:44.0 & 14:30:32 & 49.49 & -0.39 & 241.8 & 1.3 & 0.47 & 0.02 & 44.2 & 105.7 & 0.9 & 105.7 & 0.9 & Sagittarius$^h$ & 7$\pm$1.5(22)\\
IRAS 20126+4104$^{21}$ & 20:14:29.4 & 41:13:34 & 78.13 & 3.62 & 16.6 & 1.6 & 0.49 & 0.37 & 33.0 & 89.2 & 21.4 & 89.2 & 21.4 & Local & 1.5$\pm$0.5(23)\\
W75 N$^{22}$ & 20:38:36.4 & 42:37:35 & 81.87 & 0.78 & 158.8 & 4.1 & 0.22 & 0.09 & 80.0 & 132.3 & 11.0 & 61.1 & 1.0 & Local$^i$ & 2-3(24)(25)\\
DR21$^{22}$ & 20:39:01.0 & 42:19:31 & 81.68 & 0.54 & 131.0 & 3.2 & 1.05 & 0.03 & 8.0 & 60.2 & 0.8 &   &   & Local$^i$ & 2-3(24)(25)\\
\enddata
\tablecomments{Angles given represent the E-field. 8\,kpc was quoted for objects in Galactic Center.
\\ a. Numeric exponents denote the ``complex'' groupings for each source.
\\ b. $\langle\theta_G\rangle$ and uncertainties for the object's associated complex. Values listed only once for each complex.	
\\ c. Distances are from (1) \citealt{xu06}; (2) \citealt{deZ99}; (3) \citealt{ken94}; (4) \citealt{men07}; (5) \citealt{ant82}; (6) \citealt{her76}; (7) \citealt{rod82}; (8) \citealt{mof79}; (9) \citealt{par02}; (10) \citealt{wal56}; (11) \citealt{mam08}; (12) \citealt{ima07}; (13) \citealt{lau97}; (14) \citealt{nec78}; (15) distances for objects in the Galactic center were all taken to be 8 \,kpc; (16) \citealt{hel07}; (17) \citealt{dam85}; (18) \citealt{pov07}; (19) \citealt{wil70}; (20) \citealt{win83}; (21) \citealt{gwi92}; (22) \citealt{gen81}; (23) \citealt{shi08}; (24) \citealt{cam82}; and  (25) \citealt{ode93}.
\\ d. Kinematic distance is reported to be 10.0\,kpc by \citep{wal97}, but 8\,kpc was adopted here.
\\ e. Uncertainties may place these objects in the Near 3\,kpc Arm.
\\ f. Uncertainties may place this object in the Far 3\,kpc Arm or the Long Bar.
\\ g. Uncertainties may place this object in the Outer Arm.
\\ h. Uncertainties may place this object in the Perseus Arm.
\\ i. There is much discussion on the exact distance to these objects. Uncertainties may place these objects just outside Local Arm towards the Perseus Arm.
}
\end{deluxetable}

\begin{deluxetable}{lcccccc}
\tablewidth{0pt}
\tablehead{\colhead{$R^2$ Comparison\tablenotemark{a}} & \colhead{All Data} & \colhead{Complexes} & \colhead{$-9< l < 82$} & \colhead{$196 < l < 214$} & \colhead{Gal.\ center} & \colhead{Local Arm}}
\startdata
$\langle\theta_G\rangle$ vs.\ $l$ & $<$0.01 & 0.06 & $<$0.01 & $<$0.01 & 0.12 & $<$0.01 \\
$\langle\theta_G\rangle$ vs.\ $b$ & $<$0.01 & 0.13 & $<$0.01 & 0.04 & 0.05 & 0.04 \\
$\langle P \rangle$ vs.\ $l$ & 0.12 & 0.23 & 0.06 & 0.02 & 0.08 & 0.11 \\
$\langle P \rangle$ vs.\ $b$ & 0.09 & 0.07 & $<$0.01 & $<$0.01 & $<$0.01 & 0.08 \\
$\langle I \rangle$ vs.\ $l$ & 0.16 & 0.23 & 0.02 & 0.18 & 0.36\tablenotemark{b} & $<$0.01 \\
$\langle I \rangle$ vs.\ $b$ & 0.02 & $<$0.01 & 0.10 & 0.08 & $<$0.01 & 0.02 \\
$\langle\theta_G\rangle$ vs.\ $\langle I \rangle$ & 0.02 & 0.02 & 0.03 & 0.03 & 0.02 & 0.03 \\
$\langle\theta_G\rangle$ vs.\ $I_\mathrm{tot}$ & 0.02 & 0.06 & 0.01 & 0.09 & 0.02 & 0.05 \\
$\langle P \rangle$ vs.\ $\langle I \rangle$ & 0.06 & 0.04 & 0.03 & 0.04 & 0.13 & 0.01 \\
$\langle P \rangle$ vs.\ $I_\mathrm{tot}$ & $<$0.01 & $<$0.01 & 0.02 & 0.01 & 0.12 & 0.03 \\
\enddata
\tablecomments{For the reported comparisons, $\langle\theta_G\rangle$'s were kept between $0\arcdeg$--$180^\circ$. Values of Galactic coordinate $l$ that were above $350^\circ$ degrees were made negative. }
\tablenotetext{a}{Values for each comparison is the Pearson product-moment correlation coefficient, $R^2$. }
\tablenotetext{b}{Sgr B2 has a very large intensity that is a definite outlier. Removing it causes $R^2$ to decrease to 0.04.} 	
\end{deluxetable}

\subsection{Polarization and Position Angle vs.\ Galactic Coordinates}

Figure 1 shows four scatter plots for the 52 Galactic Hertz datasets: polarization
and $\langle\theta_G\rangle$ vs.\ both Galactic coordinates. There
is no correlation in these graphs. As expected, in plots involving the
Galactic latitude $b$, most objects cluster around $|b|\sim 0^\circ$, as clouds lie within the Galactic plane.  Those with values of $b$ that are not near $0^\circ$ are all objects in the local arm (determination of spiral arm is discussed in \S 3.2).  Note that polarization measurements with
$\sigma_{\langle P \rangle}$ $\geq$ $\langle$P$\rangle$ are not significant.

\begin{figure}
\includegraphics[width=1\textwidth, angle=0]{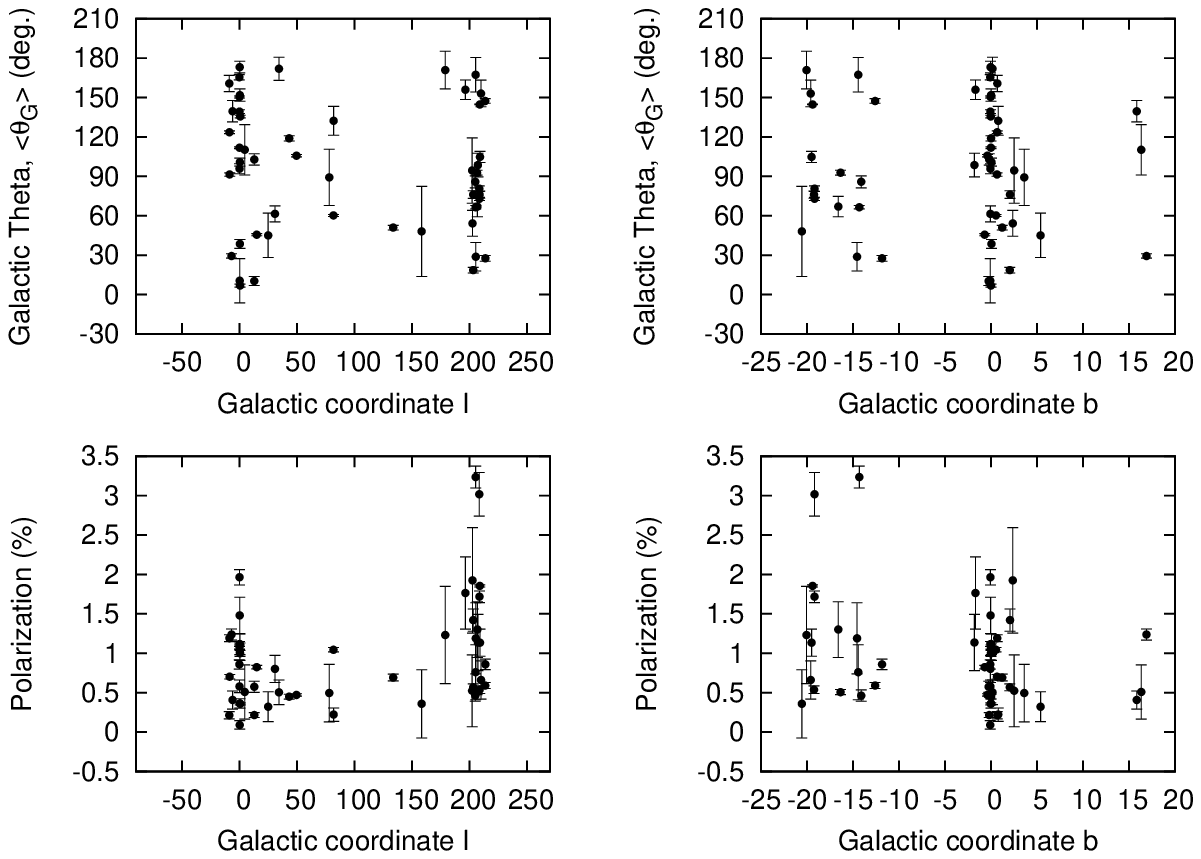}
\caption{$\langle\theta_G\rangle$ and polarization vs.\ Galactic coordinates, all data. Error bars are $1\sigma$.}
\end{figure}

There is a sufficient gap in data points between $l\simeq82^\circ$ and $l\simeq196^\circ$ as well
as between $l\simeq214^\circ$ and $l\simeq351^\circ$. Binning data
into the two prevalent set of points (i.e., around $l\simeq0^\circ$
and $l\simeq200^\circ$) still does not yield a correlation. Furthermore,
analysis of $\langle\theta_G\rangle$ and $\langle P\rangle$ vs.\
$\langle I\rangle$ and $I_\mathrm{tot}$ was made in attempt to see if
brighter objects affect $\langle\theta_G\rangle$ and $\langle P\rangle$.
No relationship was found here either.

Data were also separated into 22 complexes and analyzed in the same way as above. Figure~2 shows the same plots as Figure 1, but with the datasets reduced to 22 complexes. Again, no correlations were found in any of the relationships.
 
\begin{figure}
\includegraphics[width=1\textwidth, angle=0]{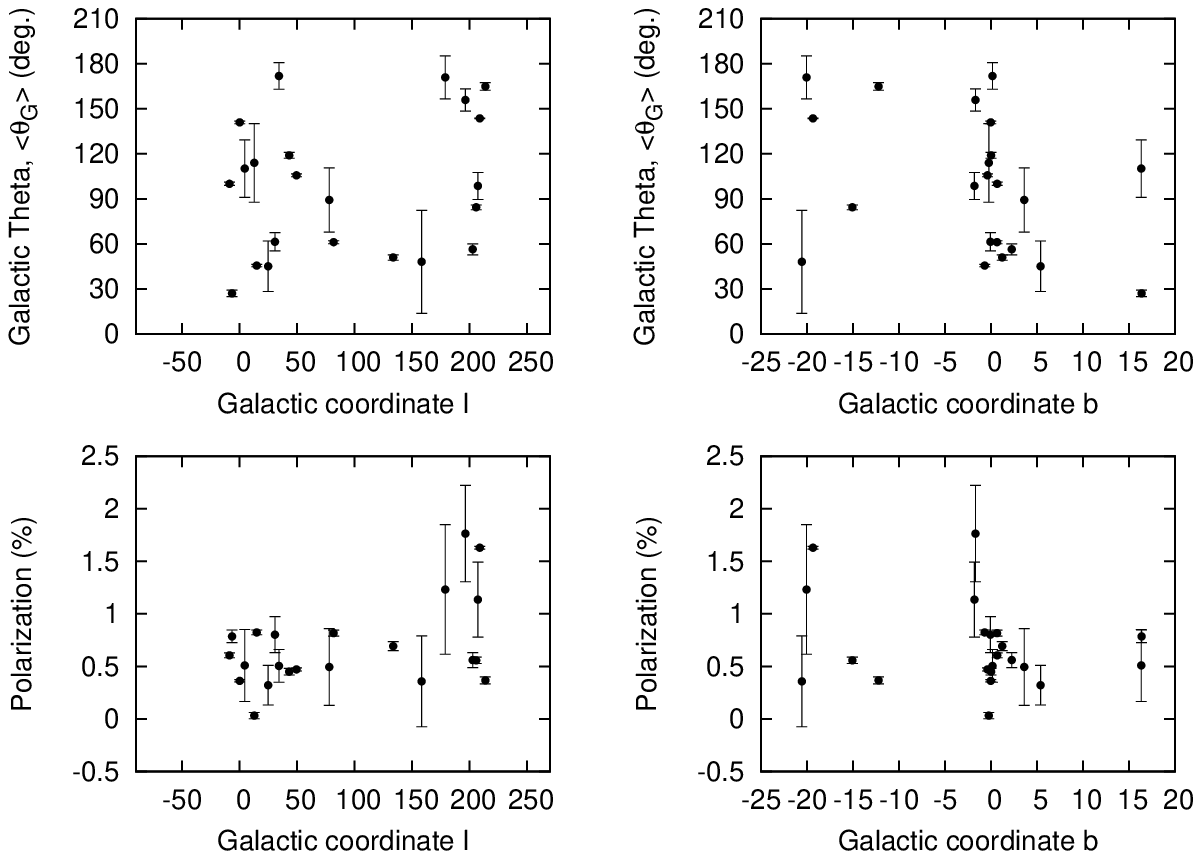}
\caption{$\langle$$\theta_G$$\rangle$ and polarization vs.\ Galactic coordinates, complexes. Error bars are $1\sigma$.}
\end{figure}

\subsection{Binning Galactic {\it l} according to spiral arms}
The final attempt to find correlations between location and the resulting large-scale angle averages was to plot values based on the Galactic arm in which each object is located. In order to find the corresponding object and its relevant Galactic arm, knowledge of the distance to each object is required. We used SIMBAD to confirm that the distances found in the literature for each object (see Table 1) corresponded to the correct objects. By using a graphical grid of the Milky Way \citep{rei09} and basic trigonometry from distances and Galactic coordinates $l$ and $b$, the correct spiral arms were ascertained. Table 3 shows the approximate distribution of data contained within each spiral arm, while Table 1 indicates the corresponding spiral arm for each object. The spiral ``arms'' containing the most data points are the Local Arm and the Galactic center. The arms with insufficient amounts of data were not thoroughly analyzed, though they seemed relatively random as well.


\begin{deluxetable}{cc}
\tablewidth{0pt}
\tablehead{\colhead{Spiral Arm} & \colhead{Number of Hertz Clouds}}
\startdata
Perseus & 4\\
Local & 27\\
Sagittarius & 5\\
Scutum-Crux & 4\\
Galactic center & 12\\
\enddata
\end{deluxetable}

Once again, the same analysis as in Section 3.1 showed no significant correlations within spiral arm bins.

\subsection{Area of avoidance}

The Galactic magnetic field is thought to follow the spiral arms. Along the line of sight, the electric field of the radiation will be orthogonal to the magnetic field, i.e., pointing towards the north Galactic pole (NGP). Since $\langle\theta_G\rangle$ is measured from the NGP, it is expected that a Galactic effect on the polarization would tend to point towards $0^\circ$ and avoid $90^\circ$ (i.e., orthogonal to the Milky Way's magnetic field). However, this trend is \emph{not}\/ seen in Figure 3. For both large and small polarization percentages, many objects lie very near or on top of this ``area of avoidance''.

\begin{figure}
\includegraphics[width=1\textwidth]{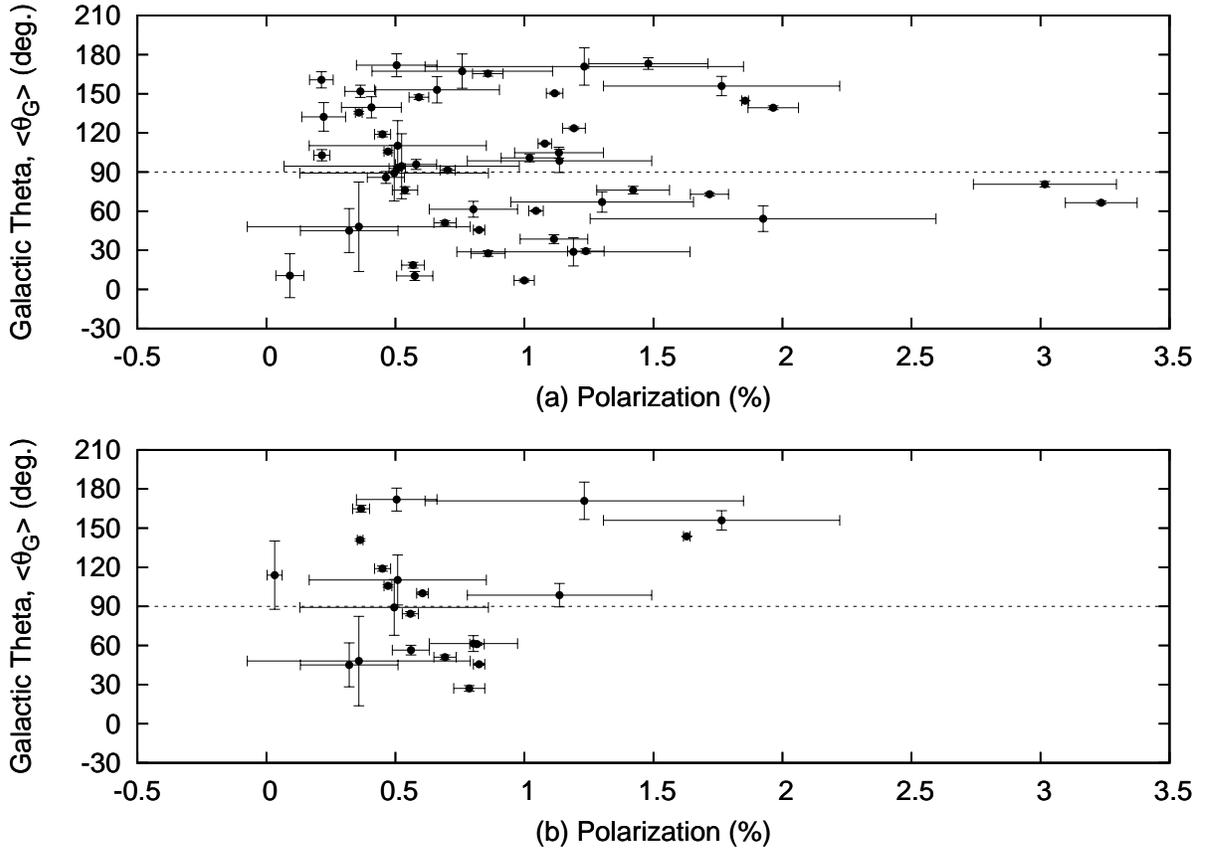}
\caption{$\langle\theta_G\rangle$ vs.\ Polarization. (\emph{a}) all data (\emph{b}) complexes. Error bars are $1\sigma$. The dashed line indicates the expected Galactic angle of avoidance at $\langle\theta_G\rangle = 90^\circ$. It does not seem that the data try to avoid this particular angle.}
\end{figure}

\subsection{Distribution of polarization angles}
The Heiles catalog \citep{hei00} has optical polarization measurements toward over 9000 stars. Those with $P>0.2$\% and distance $>$140\,pc are seen in black in Figures 4. The inferred magnetic field vectors are shown in white for the 52 Hertz clouds in Figure 4a and the 22 Hertz complexes in Figure 4b. For the most part, the submillimeter angles look randomly distributed; no obvious correlation can be seen with respect to the Galactic plane. While the B-vectors of the optical data tend to be parallel to the disk of the Galaxy, the submillimeter dust polarization in star-forming regions appears to have no directional preference. A detailed comparison of Hertz data to spatially co-located Heiles data is discussed in \cite{li09}.

It also should be noted that polarimetric maps from WMAP, which are likely dominated by dust at 94 GHz, show dust magnetic field lines parallel to the Galactic magnetic field \citep{hin09}.

\begin{landscape}
\begin{figure*}[h]
\includegraphics[width=1.3\textwidth]{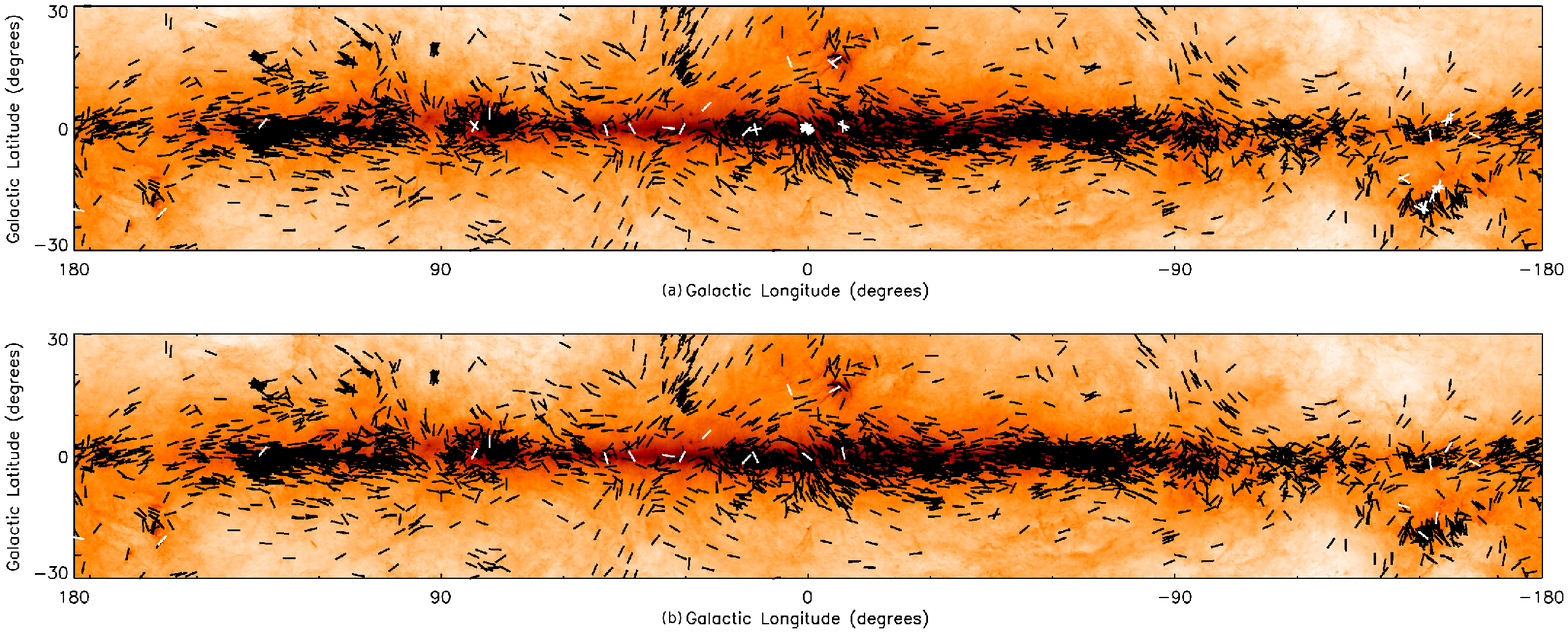}
\caption{Inferred magnetic field orientations plotted in Galactic
 coordinates. Data from Hertz datasets (in white, with (\emph{a}) all data; and (\emph{b}) complexes) are shown along with background starlight polarization from the \citet{hei00} catalog (black lines, $P>0.2$\%, distance $>140$\,pc).  The grayscale shows the IRAS $100\,\micron$ intensity data on a logarithmic scale \citep{miv05}.}
\end{figure*}
\end{landscape}


\begin{figure}
\includegraphics[width=0.8\textwidth]{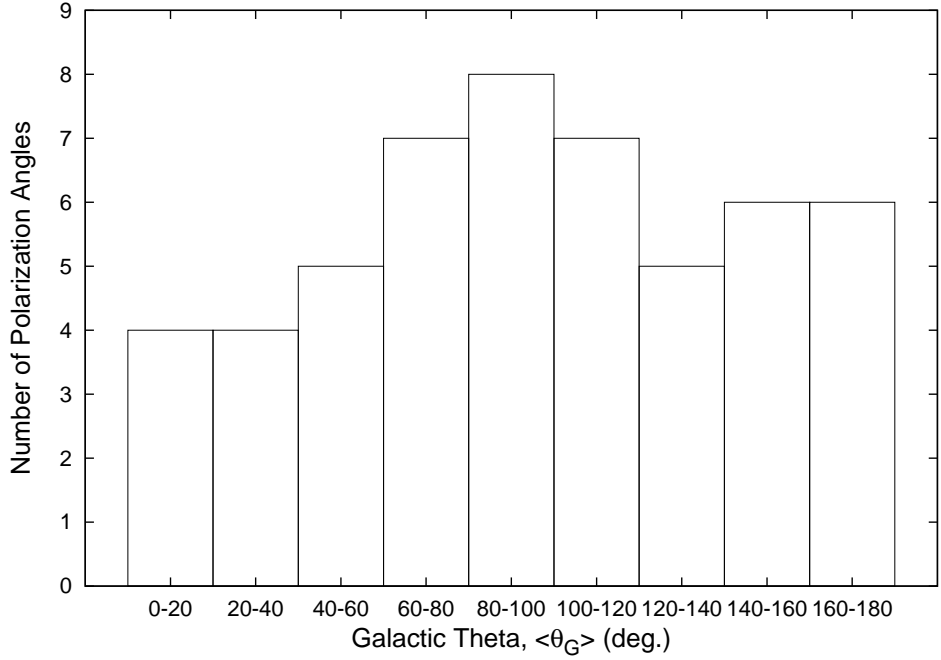}
\caption{Histogram of polarization angles for the 52 datasets with 20$^\circ$-wide bins.}
\end{figure}

A histogram of $\langle\theta_G\rangle$ for all 52 datasets is seen in Figure 5 and on inspection appears consistent with a uniform distribution. A uniform distribution for angles between 0$^\circ$ and 180$^\circ$ has standard deviation of 52.0$^\circ$. Table 4 summarizes results that test whether the sources and their complexes are uniform. The standard deviations are close to 52.0$^\circ$. Rows 2 and 4 are perhaps more valid because these remove datasets that appeared ``random'' as discussed in \S 2.2. A KS-test was used to compare each distribution to a uniform distribution. The p-values of these tests are shown in Table 4; high p-value indicates consistency with a uniform distribution. Since all p-values are high, there is no preferred direction in the sky.

\begin{deluxetable}{lccc}
\tablewidth{0pt}
\tablehead{\colhead{Data Analyzed} & \colhead{$\sigma_\mathrm{dist}$ (deg.)} & \colhead{KS-test P-value}}
\startdata
All Datasets (52) & 48.6 & 0.69\\
KS-test accepted datasets (42) & 47.7 & 0.69\\
All complexes (22) & 45.5 & 0.28\\
KS-test accepted complexes (17) & 47.5 & 0.51\\
\enddata
\tablecomments{The value in parentheses indicates the number of datasets.}
\end{deluxetable}

As a comparison to the distribution of Hertz angles, a histogram of the Heiles data with $P>0.2$\% and distance $>140$\,pc is shown in Figure 6. These cuts were chosen because all Hertz datasets except one (M+0.11-0.08 at $P=0.09$\%) fit this criteria. There is overwhelming evidence that the data are a non-uniform distribution centered around a polarization position angle of 0$^\circ$. Plotting Heiles angles without cuts show similar graphs. Additional filtering was done based on larger $P$ and extinction cuts ($E(B-V)$ in the Heiles database) because these areas generally probe denser regions. These cuts tighten the distribution around 0$^\circ$.

\begin{figure}
\includegraphics[width=0.8\textwidth]{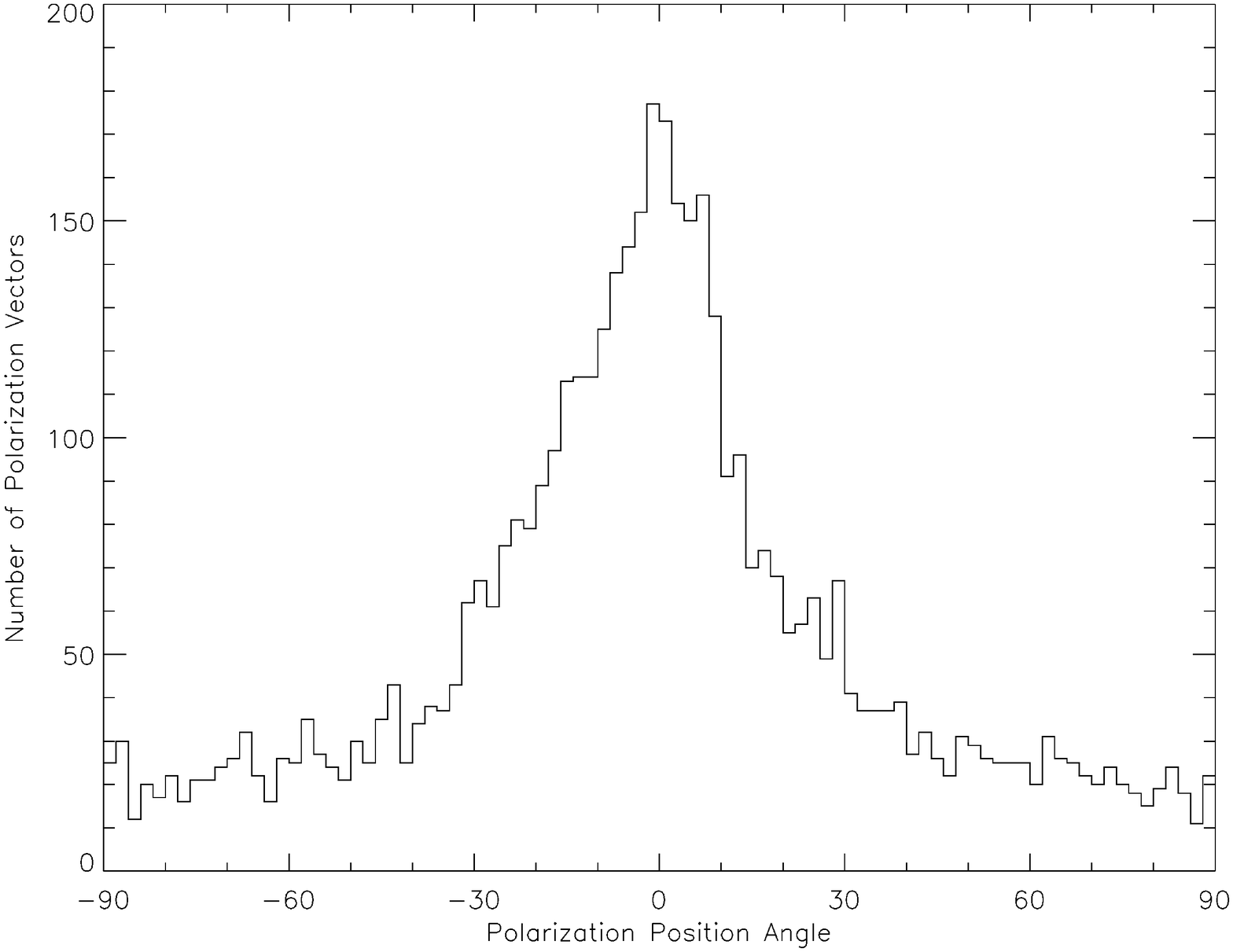}
\caption{Histogram of Heiles polarization angles (with respect to Galactic coordinates) with $P>0.2$\% and distance $>140$\,pc.}
\end{figure}

\section{Discussion and Conclusion}\label{discussion}

We have used Hertz \citep{dot10} polarization measurements in dense, star-forming clouds, to investigate whether the ordered component of the magnetic field in these clouds is correlated with location in the Galaxy. 
In order to obtain a single polarization percentage, angle, and intensity for each of the 52 Hertz objects (as well as the 22 complexes), the polarization information for each beam was combined into a large-scale average. With each object represented by these values, correlations with respect to location in the Galaxy (both by Galactic coordinates and spiral arm locations) were investigated. 

There are three primary results for this paper:
\begin{itemize}
\item A meaningful polarization angle can be determined for most objects and complexes.
\item No evidence was found in our data for a correlation between the polarization angle and location within the galaxy.
\item The polarization angle for an object or a complex on the sky is consistent with a random distribution.
\end{itemize}


The fact that a meaningful mean direction can be identified for the
magnetic field in most objects implies the existence of an ordered,
large scale component of the field {\em within} the dense,
star-forming clouds we have studied. This is consistent with the
continuity of magnetic field direction on different scales within
these clouds, discussed by \cite{li09}. 
However, since these are star-forming clouds, feedback processes from newly formed stars can
generate appreciable scatter in the magnetic field directions within
each object. This may be, in part, the source of the observed
angle dispersion within each dataset. 


We have found that there is no significant evidence for the existence of any correlation between mean polarization angle and location, which is consistent with the results by \citet{gle99} (see \S1). The fact that objects do not seem to avoid polarization angles aligned with the Galaxy's magnetic field implies that the polarization angles detected are almost entirely created by the analyzed object rather than a large-scale, external field. This suggests that complexes as a whole may become their own dynamical system that is separate from the Galaxy. The results in this paper imply that cloud cores usually have a meaningful net field (which may correlate with other cores in the same complex, \citep{li09}) that has no preferred direction within the Galaxy, yet are embedded in a diffuse medium in an ordered, Galactic large-scale field. 

\citet{fis03} did a similar analysis as this paper, but with a different methodology. They analyzed the line-of-sight magnetic fields through Zeeman splitting of OH masers in massive star-forming regions; on sub-kiloparsec scales (about 0.5\,kpc), two sources often had opposite line-of-sight field directions, suggesting multiple cores in a complex tangle the magnetic fields. In some areas of the Galaxy, \citeauthor{fis03} found some line-of-sight field alignment in parts of the Sagittarius Arm and Norma Arm on scales of about 2\,kpc. Still, they also found no evidence for correlations of magnetic field directions in star-forming regions with the Galactic field or with the spiral arms on larger scales.
 
The cloud formation process involves instabilities on Galactic scales \citep{shettyo08,mkc09,tt09}, which are responsible for the accumulation of enough mass to form the clouds. At the same time, these instabilities generate turbulence in the ISM of the Galaxy. The cloud magnetic field is thus expected to decouple from the Galactic field during the cloud formation process. Stellar feedback is an additional mechanism driving the cloud magnetic field away from alignment with the Galactic field direction. These effects are likely responsible for the dichotomy between the arm/interarm regions in terms of the ratio of strengths between the ordered and tangled components of the magnetic field observed in external galaxies \citep{bec05}.


\acknowledgments
\emph{Acknowledgments:} Part of this work was carried out at the Jet Propulsion Laboratory, California Institute of Technology, under a contract with the National Aeronautics and Space Administration. This research has made use of the SIMBAD database, operated at CDS, Strasbourg, France. L. W. L. acknowledges support from from the National Science Foundation under grant No. AST-07-09206. We would also like to acknowledge Richard M. Crutcher for extensive discussions.


\appendix
\section{The Equatorial to Galactic Angle Transformation}
Our goal is to convert the equatorial position angles (measured east from the north celestial pole (NCP)) to Galactic position angles (measured towards increasing longitude from the north Galactic pole (NGP)). 
The location of the NCP (B1950) is exactly $l_n = 123^\circ$ and $b_n = 27.4^\circ$. By definition the NGP is at $b_\mathrm{ngp} = 90^\circ$.

\begin{figure}
\epsscale{.50}
\plotone{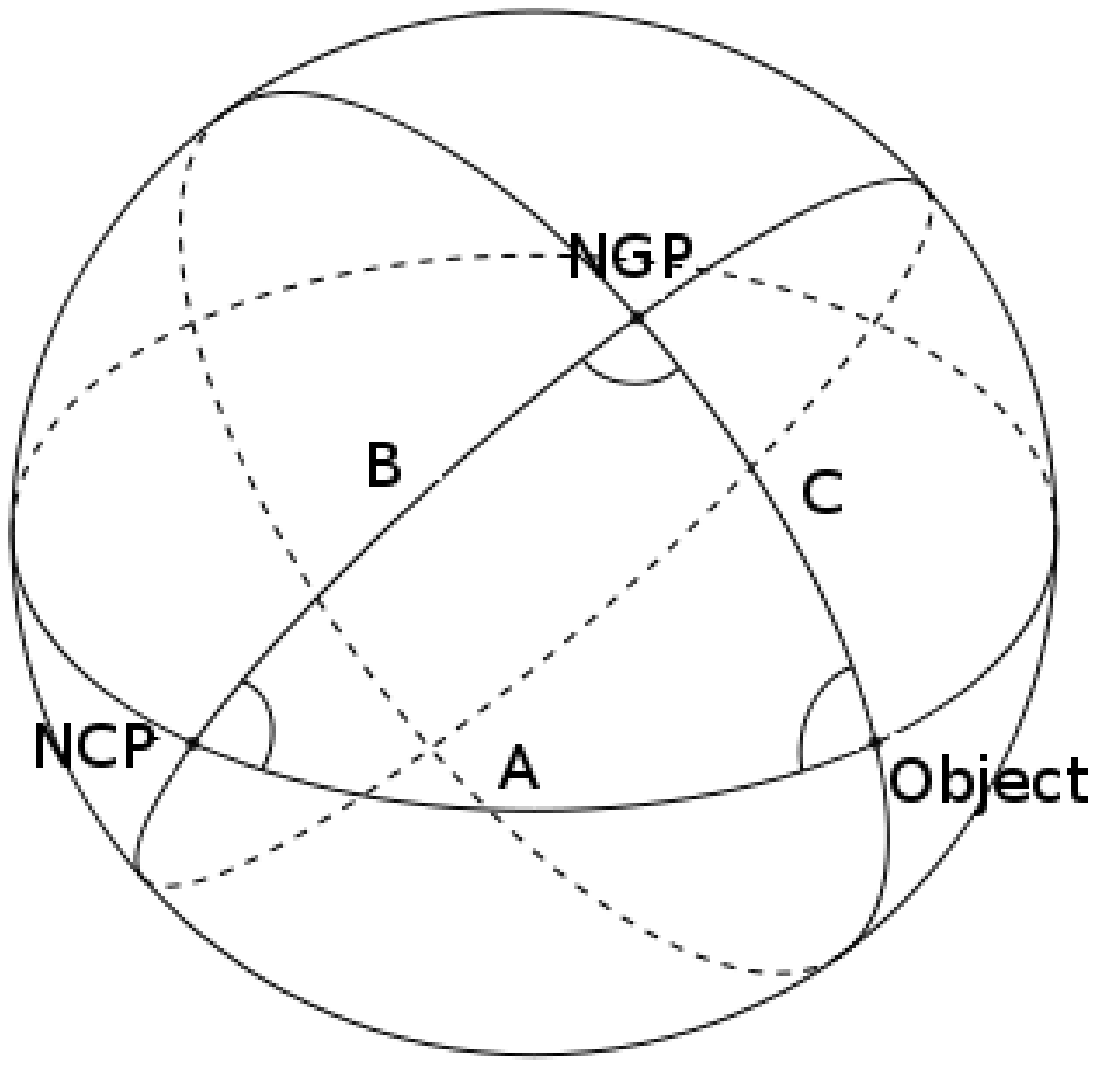}
\caption{The spherical trigonometry required to convert equatorial position angles $\theta$ into Galactic position angles $\langle\theta_G\rangle$.}
\end{figure}

 Let $l_1$ and $b_1$ be the location of an individual object. 
The rotation of the equatorial angle $\theta$ into a Galactic angle $\theta_G$ is simply the angle created at the object by lines from the NGP and NCP, as seen in Figure 7, $\angle$Object ($\angle$O). We see the angle of arc B is always $b_\mathrm{ngp} - b_n = 90^\circ-27.4^\circ = 62.6^\circ$. Similarly, arc C is $90^\circ - b_1$, but it is unimportant with knowledge of the declination. $\angle$NGP is $l_n-l_1$. From the law of cosines, we could use B and C to find arc A, but this is simply $90^\circ$ minus the B1950 declination, $\delta$. Therefore,  $A = 90^\circ-\delta$. Now, from the law of sines we can derive an equation for $\angle$O.

\begin{equation}
 \frac{\sin\angle O}{\sin B} = \frac{\sin\angle \mathrm{NGP}}{\sin A}
\end{equation}

\begin{equation}
 \angle O = \arcsin\left[\frac{\sin(123-l_1)\sin62.6^\circ}{\sin(90^\circ-\delta)}\right]
\end{equation}

\begin{equation}
  \langle\theta_G\rangle=\angle O + \langle\theta\rangle
\end{equation}

$\angle$O will either be positive or negative and is added to $\langle\theta\rangle$ to get the Galactic polarization angle, $\langle\theta_G\rangle$. Since polarization vectors are ``headless'' (have no preferred direction), $180^\circ$ is added or subtracted to the final value to make $\langle\theta_G\rangle$ between $0^\circ$ and $180^\circ$. 
Databases such as \citep{mat70} and \citep{kla77} have used this standard in the past, and this rotation method agrees with those databases to the tenth of a degree.



\end{document}